\documentclass[modern]{aastex62}
\usepackage{amssymb}
\usepackage{amsmath}
\usepackage{xspace}
\usepackage{pifont}

\newcommand{\mearth}{$M_\oplus$\xspace}
\newcommand{\rearth}{$R_\oplus$\xspace}

\newcommand{\permean}{\ensuremath{\left< P \right>}\xspace}

\graphicspath{{./}{figures/}}

\received{September 16, 2019}
\revised{\today}
\accepted{}
\submitjournal{AAS Journals}

\shorttitle{A Kepler multiplanet system progenitor}
\shortauthors{David et al.}

\begin{document}

\title{Four newborn planets transiting the young solar analog V1298 Tau}

\correspondingauthor{Trevor J. David}
\email{trevorjdavid@gmail.com}

\author[0000-0001-6534-6246]{Trevor J.\ David}
\affil{Jet Propulsion Laboratory, California Institute of Technology, 4800 Oak Grove Drive, Pasadena, CA 91109, USA}

\author[0000-0003-0967-2893]{Erik A.\ Petigura}
\affil{Department of Physics and Astronomy, University of California, Los Angeles, CA 90095, USA}

\author[0000-0002-0296-3826]{Rodrigo Luger}
\affil{Center for Computational Astrophysics, Flatiron Institute, New York, NY 10010, USA}

\author[0000-0002-9328-5652]{Daniel Foreman-Mackey}
\affil{Center for Computational Astrophysics, Flatiron Institute, New York, NY 10010, USA}

\author[0000-0002-4881-3620]{John~H.~Livingston}
\affil{Department of Astronomy, University of Tokyo, 7-3-1 Hongo, Bunkyo-ku, Tokyo 113-0033, Japan}

\author[0000-0003-2008-1488]{Eric E.\ Mamajek}
\affil{Jet Propulsion Laboratory, California Institute of Technology, 4800 Oak Grove Drive, Pasadena, CA 91109, USA}
\affil{Department of Physics \& Astronomy, University of Rochester, Rochester, NY 14627, USA}

\author{Lynne A.\ Hillenbrand}
\affil{Department of Astronomy, California Institute of Technology, Pasadena, CA 91125, USA}

\begin{abstract}
Exoplanets orbiting pre-main sequence stars are laboratories for studying planet evolution processes, including atmospheric loss, orbital migration, and radiative cooling. V1298 Tau, a young solar analog with an age of 23 $\pm$ 4 Myr, is one such laboratory. The star is already known to host a Jupiter-sized planet on a 24 day orbit. Here, we report the discovery of three additional planets --- all between the size of Neptune and Saturn --- based on our analysis of \textit{K2} Campaign 4 photometry. Planets c and d have sizes of 5.6 and 6.4 \rearth, respectively and with orbital periods of 8.25 and 12.40 days reside 0.25\% outside of the nominal 3:2 mean-motion resonance. Planet e is 8.7 \rearth in size but only transited once in the \textit{K2} time series and thus has a period longer than 36 days, but likely shorter than 223 days. The V1298 Tau system may be a precursor to the compact multiplanet systems found to be common by the \textit{Kepler} mission. However, the large planet sizes stand in sharp contrast to the vast majority of \textit{Kepler} multis which have planets smaller than 3 \rearth. Simple dynamical arguments suggest total masses of $<$28 \mearth and $<$120 \mearth for the c-d and d-b planet pairs, respectively. The implied low masses suggest that the planets may still be radiatively cooling and contracting, and perhaps losing atmosphere. The V1298 Tau system offers rich prospects for further follow-up including atmospheric characterization by transmission or eclipse spectroscopy, dynamical characterization through transit-timing variations, and measurements of planet masses and obliquities by radial velocities.
\end{abstract}

\keywords{Exoplanet astronomy -- Exoplanet evolution -- Transit photometry -- Planetary system formation -- Weak-line T Tauri stars -- Young star clusters}

\section{Introduction} \label{sec:intro}
Compact multiplanet systems are one of the signature discoveries of NASA's \textit{Kepler} mission \citep{Borucki:etal:2010, Steffen:etal:2010, Lissauer:etal:2011}. These planetary systems are ubiquitous in the Galaxy yet much about their origins remains a mystery. In general, the orbits of planets in \textit{Kepler} multiplanet systems are nearly circular \citep{Hadden:Lithwick:2014, VanEylen:Albrecht:2015, Xie:etal:2016} and coplanar \citep{Tremaine:Dong:2012, Fang:Margot:2012, Fabrycky:etal:2014} with relatively low obliquities \citep{SanchisOjeda2012, Hirano:etal:2012, Chaplin:etal:2013, Albrecht:etal:2013, Morton:Winn:2014}. 

There is also a high degree of intra-system uniformity amongst planets in multi-transiting systems; the masses, radii, and orbital spacing of adjacent planets in a given system are more similar than planet pairs chosen at random from the overall population of multiplanet systems \citep{Lissauer:etal:2011, Millholland2017, Weiss2018}. The orbital spacings between adjacent planets in multi-transiting systems are well-described by a Rayleigh distribution, with a peak near 20 mutual Hill radii \citep{Fang:Margot:2013, Pu:Wu:2015, Dawson:etal:2016, Weiss2018}. At the small separation end of this distribution some planetary systems are on the verge of instability \citep{Deck:etal:2012, Pu:Wu:2015}. While there is a small but significant excess of planet pairs in and just outside of low-order resonances, the majority of planets in \textit{Kepler} multi-transiting systems are not near a resonance \citep{Lissauer:etal:2011}.

Planets in \textit{Kepler} multi-transiting systems are generally smaller than Neptune ($R < 4$~\rearth) and rarely accompanied by a nearby transiting Jovian planet \citep{Latham:etal:2011}. Furthermore, the radius distribution of \textit{Kepler} planets is bimodal, with a valley near 1.7~\rearth that separates small and likely rocky planets from larger ones with substantial atmospheres and preferentially wider orbits \citep{Fulton2017}. In about 2/3 of adjacent planet pairs the outer planet is larger with a size ratio that may be correlated with the difference in equilibrium temperatures \citep{Ciardi:etal:2013, Weiss2018}. Both the bimodal size distribution and size-location correlation are seen as circumstantial evidence for past atmospheric loss \citep{OwenWu2013, LopezFortney2013}.

Occasionally, multi-transiting systems contain neighboring planets with vastly different bulk densities. Such is the case for Kepler 36 b \& c, which have semi-major axes that differ by 10\% but densities that differ by nearly an order of magnitude \citep{Carter:etal:2012}. In systems where small, rocky planets are found on orbits interior to those with substantial volatile envelopes, the density discrepancies might be explained by photo-evaporative mass loss \citep{LopezFortney2013}. In other cases, the differing densities of adjacent rocky planets might be explained by giant impacts \citep{Bonomo:etal:2019}. 

The existence of atmospheres contributing a few percent to the total planet mass for many planets in multi-transiting systems implies that envelopes are accreted before dispersal of the protoplanetary disk. However, precisely when and where the cores form is debated; formation of \textit{Kepler} planets might proceed \textit{in situ} \citep{Hansen:Murray:2012, Hansen:Murray:2013, Ikoma:Hori:2012, Chiang:Laughlin:2013, Chatterjee:Tan:2014, Lee:etal:2014, Lee:Chiang:2016}, or \textit{ex situ} followed by migration via tidal interactions with the protoplanetary disk \citep{Terquem:Papaloizou:2007, Ida:Lin:2010}. By studying the properties of planetary systems across a wide range of ages, it may be possible to constrain the initial conditions of \textit{Kepler} multiplanet systems and assess the relative likelihoods of these two formation channels. 

We previously reported the detection of a warm, Jupiter-sized planet transiting the pre-main sequence star V1298 Tau \citep{David2019}. That work also presented a statistical validation of the planet V1298 Tau b and stellar characterization which we do not reproduce here. In follow-up papers we derived a more stringent upper limit to the mass of V1298 Tau b from precision near-infrared radial velocities \citep{Beichman:etal:2019} and a revised ephemeris for that planet from \textit{Spitzer} 4.5~\micron\ transit observations (Livingston et al., in prep.). In this work we report three previously unidentified transiting planets from the \textit{K2} light curve of V1298 Tau. In \S~\ref{sec:analysis} we describe the procedures used to model the time series photometry and derive planet parameters. We consider the V1298 Tau planetary system in the context of other known multi-transiting systems in \S~\ref{sec:discussion} and present our conclusions in \S~\ref{sec:conclusions}. 

\section{Light Curve Analysis} \label{sec:analysis}

\begin{figure}
    \centering
    \includegraphics[width=\textwidth]{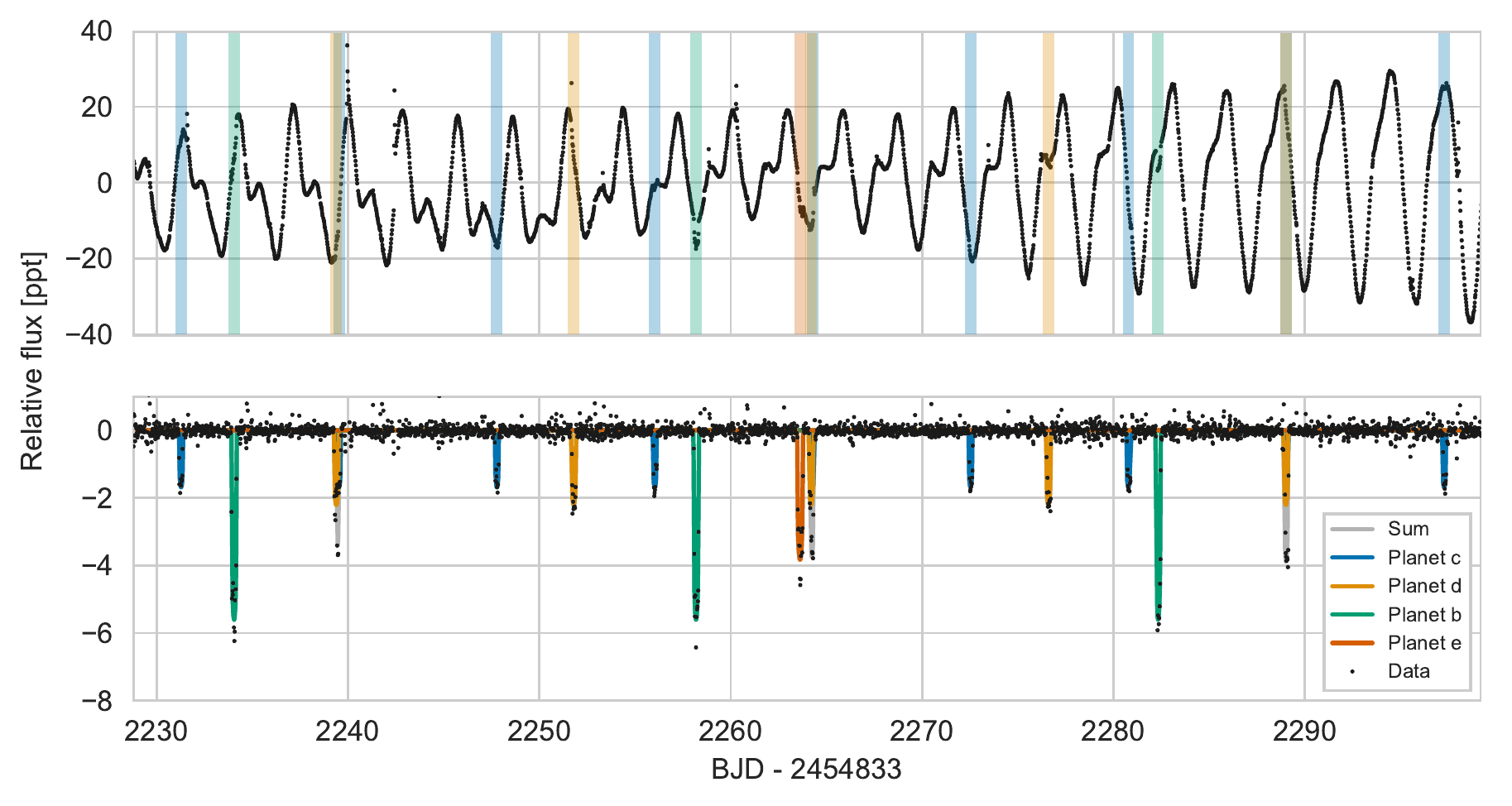}
    \includegraphics[width=\textwidth]{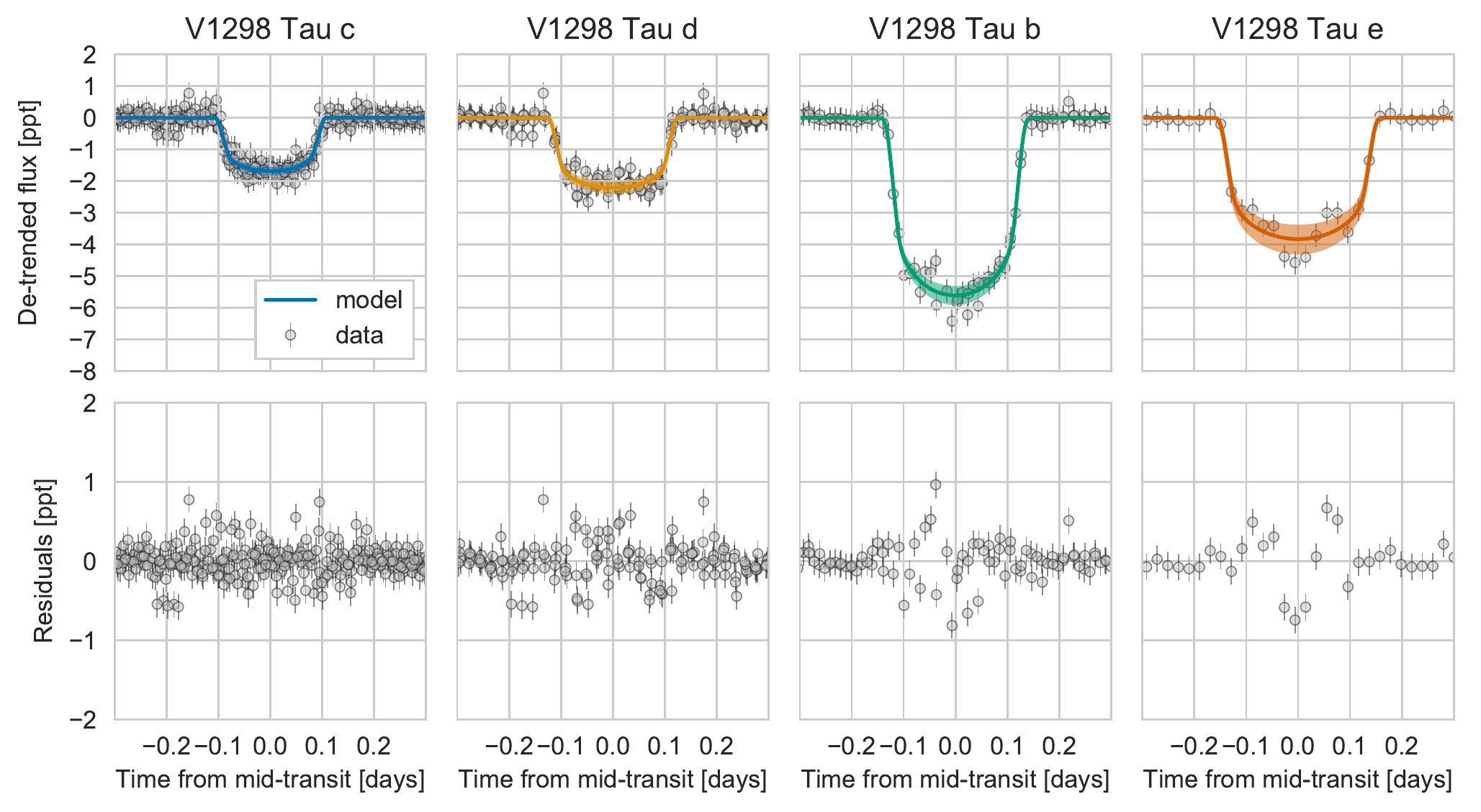}
    \caption{The full \textit{K2} light curve of V1298 Tau before (first row) and after (second row) subtracting the median GP model. Segments of the light curve including transits are shown by the shaded bands. In the second panel, the de-trended flux (data - median model) along with the median transit models for each planet are shown. In the third row, the phase-folded transits and median models are shown for each individual planet. For V1298 Tau c, data acquired during simultaneous transits are shown after subtracting the transit model of planet d, and vice versa. The shaded bands indicate the 1$\sigma$ error contours of the transit models. Residuals (data - median model) are shown in the bottom row for each planet.}
    \label{fig:lc}
\end{figure}

\begin{figure}
    \centering
    \includegraphics[width=\textwidth]{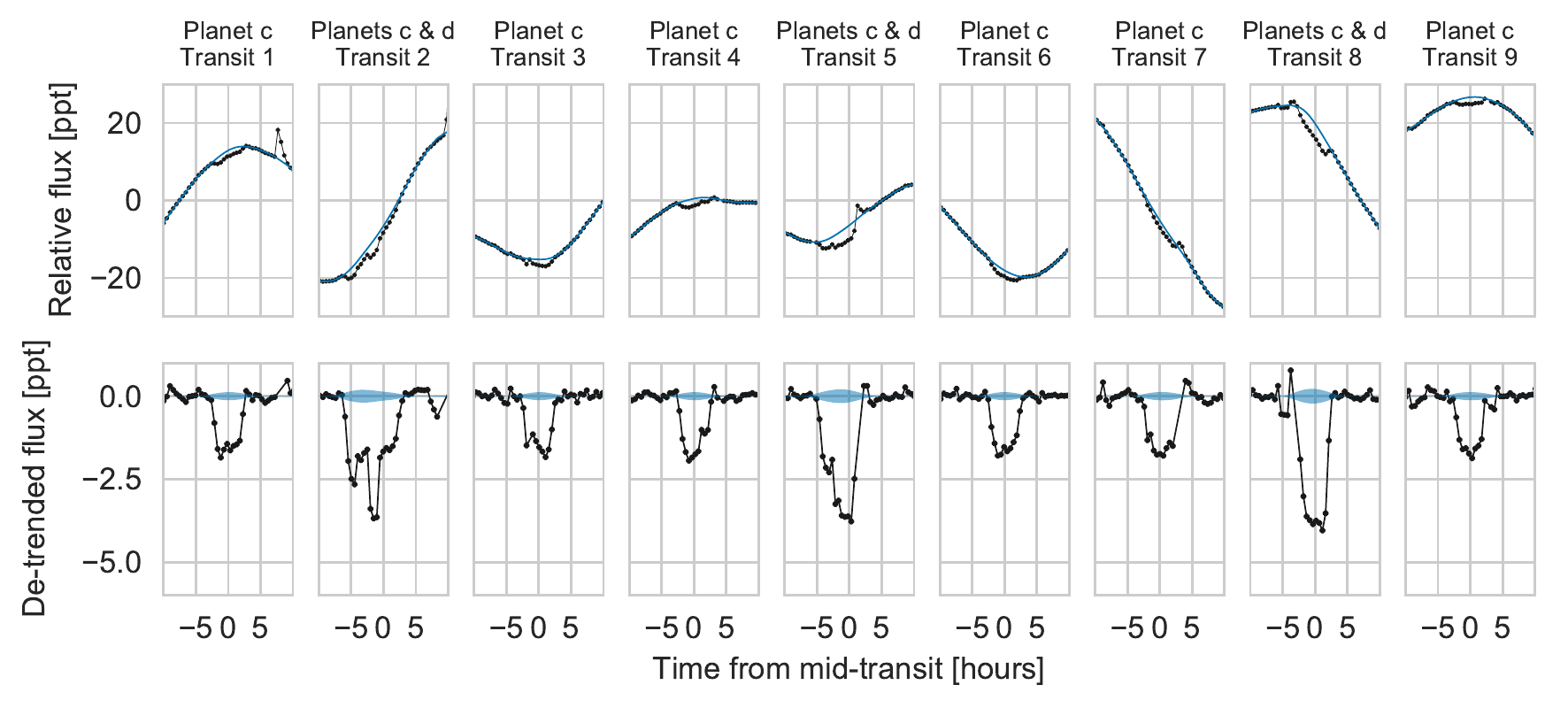}
    \includegraphics[width=\textwidth]{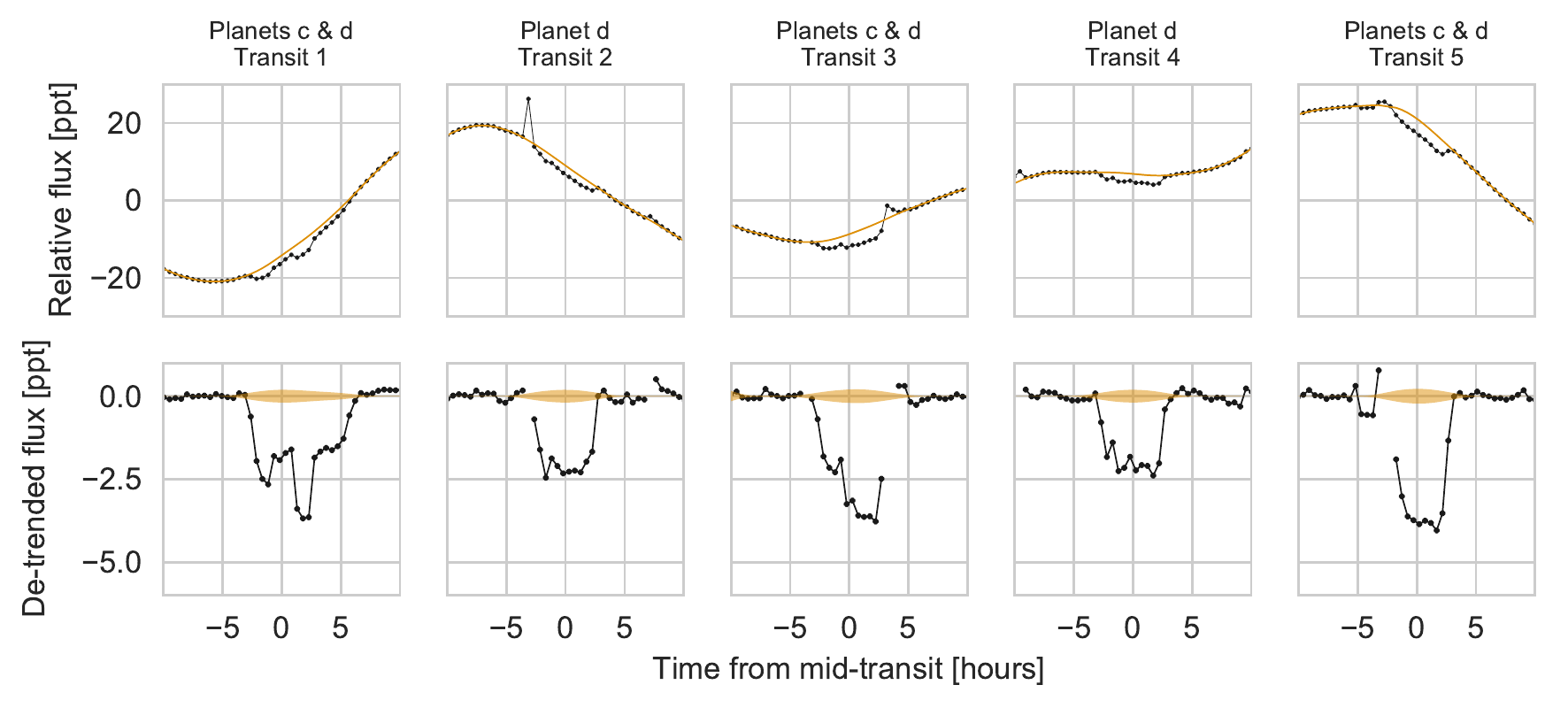}
    \includegraphics[width=\textwidth]{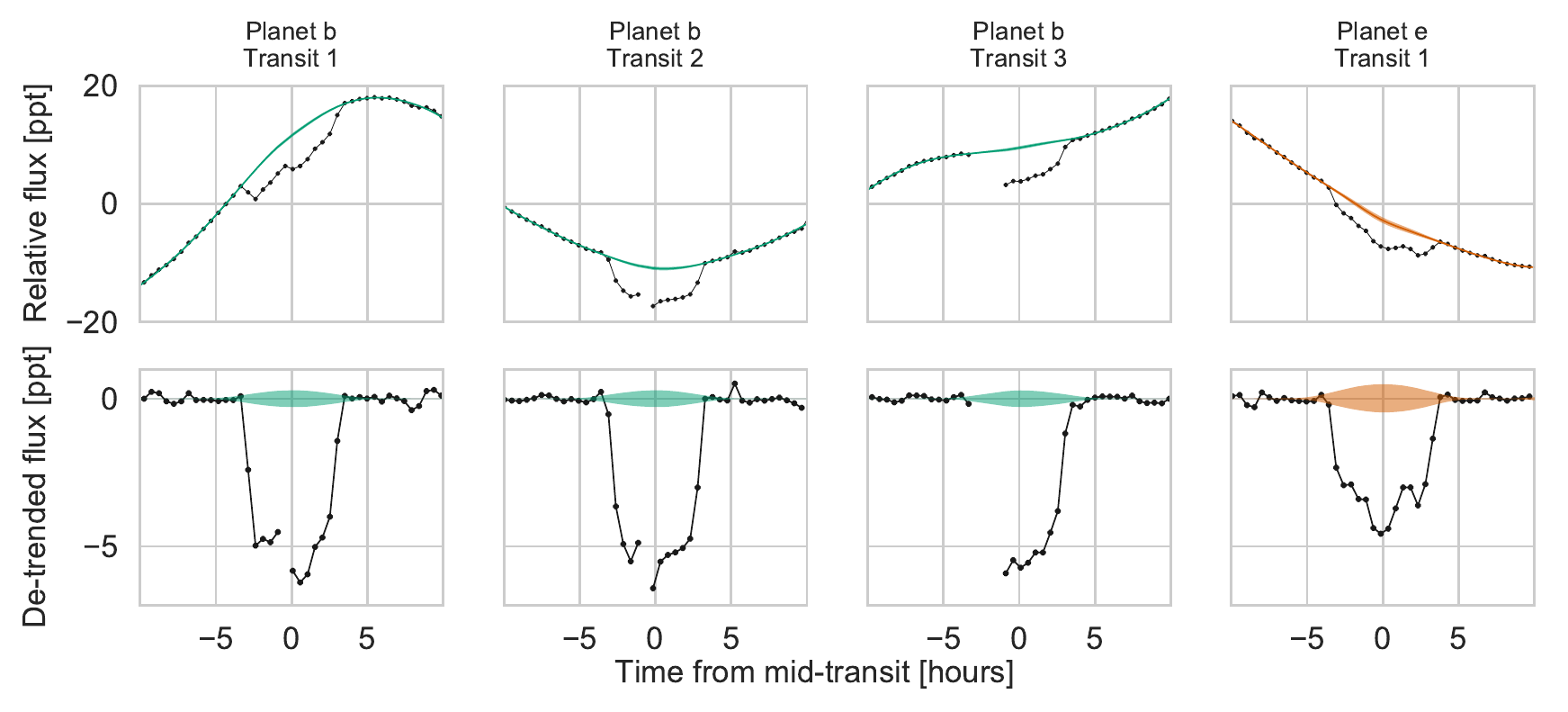}
    \caption{Individual transits of the four planets orbiting V1298 Tau. For each individual transit of each planet a 20-hour segment of the light curve is shown before (top rows) and after (bottom rows) subtracting the median GP model. The GP model and associated 1$\sigma$ error contours are shown by the shaded lines and bands, respectively. Simultaneous transits of planets c and d are indicated as such. The spacecraft did not acquire data during the ingress of the third transit of planet b due to a loss of fine pointing. Data gaps indicate missing or excluded data.}
    \label{fig:gp}
\end{figure}

\begin{figure}
    \centering
    \includegraphics[width=\textwidth]{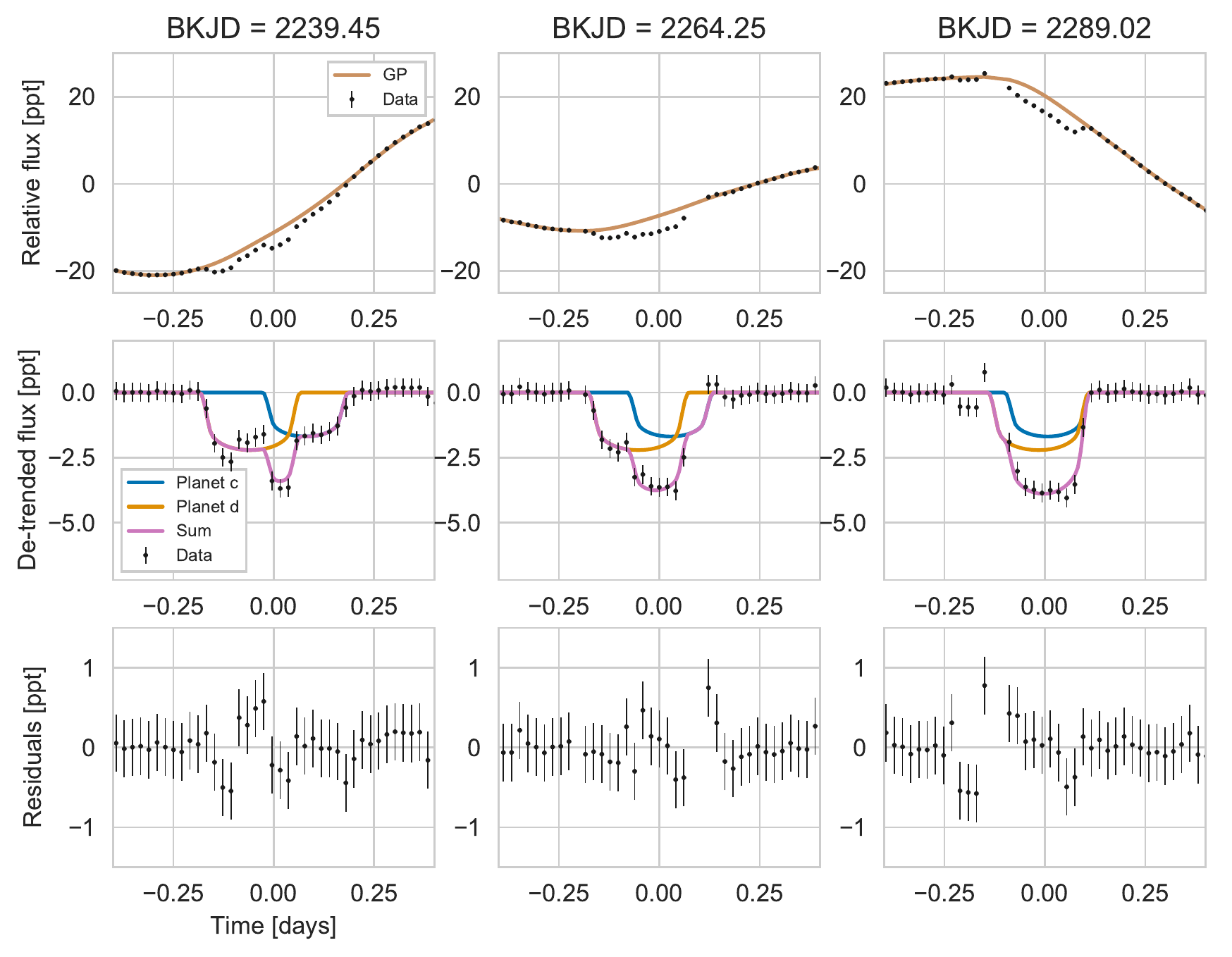}
    \caption{Simultaneous transits of planets c and d. In the top panels the \textit{K2} observations are shown as black points and the GP model is shown by a beige line. In the middle panels, the median transit models for planets c and d are shown by the blue and orange lines, respectively, as well as the sum (pink). Residuals between the data and the summed transit model are shown in the bottom panels.}
    \label{fig:simultaneoustransits}
\end{figure}

\subsection{Systematics Model}
V1298 Tau was observed by the \textit{Kepler} space telescope from 2015 February 7 to 2015 April 23 UTC during Campaign 4 of the \textit{K2} mission \citep{Howell2014}. \textit{K2} time series are affected by pointing-related systematic trends. Such systematic trends can be mostly removed through publicly available ``de-trending'' software. We adopted the \texttt{EVEREST 2.0} \citep{Luger2018} light curve, which corrects for \textit{K2} systematics using a variant of the pixel-level decorrelation method \citep{Deming2013}.

In some cases it is important to mask transits prior to the systematics de-trending procedure. Neglecting to follow this step may result in the in-transit observations influencing the systematics model and consequently distorting the transit shape. We investigated whether this step was critical to our own analysis using the \texttt{star.transitmask} attribute in \texttt{EVEREST 2.0}. We found the differences in the corrected flux time series with and without applying this mask to be $<$0.01 ppm, which is much smaller than the estimated photometric precision of 130--180 ppm.  

\subsection{Signal Detection}
The additional transits reported here were not detected in \citet{David2019} because we adopted a stellar variability model that was too flexible and accommodated the transits of the other planets. Instead, we discovered the additional transits through visual inspection of the light curve and, for the inner two planets, confirmed their periodicity with a Box-fitting Least Squares (BLS) algorithm \citep{Kovacs2002} following two approaches to de-trending: (1) a Savitzky-Golay filter using a window length of 31 cadences, a 3rd order polynomial, and 10 iterations of 3$\sigma$ outlier exclusion, and (2) a cubic spline fit with a knot spacing of 6 cadences and a transit and flare mask created through inspection. Transits of the innermost planet were also recovered using a sliding notch filter with quadratic continuum fitting (A. Rizzuto, priv. communication).

\subsection{Astrophysical and Systematic False Positives}
Astrophysical false positive scenarios were considered and argued unlikely in \citet{David2019}. We do not reproduce that analysis here, but we consider the possibility that some of the transit features noted may be related to stellar activity or unaccounted for systematic noise. None of the transit signals have periods that bear any clear relation to the stellar rotation period. The ratio between the orbital period of the $n$-th planet and the rotation period is as follows: $P_n/P_\mathrm{rot}$ = 2.88, 4.33, 8.43 for the inner three planets. We also investigated the proximity in time of individual transits to stellar flares and spacecraft thruster firings. While the shapes of some transits could plausibly be affected by the systematics correction procedure and some occur fairly close in time to flares, neither effect can satisfactorily explain all of the transits nor their periodicity. We conclude that the simplest and most probable explanation for the transit signatures is the presence of three additional planets in the system.

\subsection{Transit Model} \label{subsec:transit}
We modeled the stellar variability and transits simultaneously using a combination of the \texttt{exoplanet} \citep{exoplanet:exoplanet}, \texttt{Starry} \citep{exoplanet:luger18}, and \texttt{PyMC3} \citep{exoplanet:pymc3} packages. The transit model within the \texttt{exoplanet} package is computed with \texttt{Starry}, and in our case was described by the following parameters: the mean out-of-transit flux ($\langle f \rangle$), quadratic limb darkening coefficients ($u_1$, $u_2$), stellar mass ($M_\star/M_\odot$) and radius  ($R_\star/R_\odot$), log of the orbital period ($\ln P$), time of mid-transit ($T_0$), log of the planet radius ($\ln R_P/R_\odot$), impact parameter ($b$), eccentricity ($e$), and longitude of periastron ($\omega$). 

A quadratic limb darkening law was assumed, with the parameterization recommended by \citet{exoplanet:kipping13} for efficient and uninformative sampling of the limb darkening coefficients. We also used the \citet{Espinoza2018} parameterization of the joint radius ratio and impact parameter distribution. For all fits, a $\beta$ distribution prior was assumed for the eccentricity:

\begin{equation}
    P_\beta(e; a,b) = \frac{\Gamma(a+b)}{\Gamma(a)\Gamma(b)}e^{a-1}(1-e)^{b-1},
\end{equation}

where $\Gamma$ denotes the Gamma function and we assumed the values $a=0.867$ and $b=3.03$ as recommended by \citet{Kipping:2013ecc}.

\subsection{Stellar variability model} \label{subsec:gps}
To evaluate the likelihood of the data given the transit model we modeled the photometric variability as a Gaussian process (GP). Specifically, we used the ``Rotation'' GP kernel in \texttt{exoplanet}\footnote{\url{https://exoplanet.dfm.io/en/stable/user/api/\#exoplanet.gp.terms.RotationTerm}}, which models variability as a mixture of two stochastically driven, damped simple harmonic oscillators with undamped periods of $P_\mathrm{rot}$ and $P_\mathrm{rot}/2$. The power spectral density of this GP is:

\begin{equation}
S(\omega) = \sqrt{\frac{2}{\pi}} \frac{S_1\omega_1^4}{(\omega^2-\omega_1^2)^2 + 2\omega_1^2\omega^2} + \sqrt{\frac{2}{\pi}} \frac{S_2\omega_2^4}{(\omega^2-\omega_2^2)^2 + 2\omega_2^2\omega^2/Q^2},
\end{equation}

where,

\begin{align}
S_1 &= \frac{A}{\omega_1 Q_1},\\
S_2 &= \frac{A}{\omega_2 Q_2}\times \text{mix},\\
\omega_1 &= \frac{4\pi Q_1}{P_\mathrm{rot}\sqrt{4 Q_1^2 - 1}},\\
\omega_2 &= \frac{8\pi Q_2}{P_\mathrm{rot}\sqrt{4 Q_2^2 - 1}},\\
Q_1 &= \frac{1}{2} + Q_0 + \Delta Q,\\
Q_2 &= \frac{1}{2} + Q_0. \\ 
\end{align}

We used \texttt{celerite} \citep{ForemanMackey:etal:2017} to compute the log-likelihood of the GP,

\begin{equation}
    \ln\mathcal{L} = -{1\over2}(\textbf{f}-\textbf{m})^T K^{-1} (\textbf{f}-\textbf{m})-{1\over 2}\ln\mathrm{det}\,K,
\end{equation}

where \textbf{f} and \textbf{m} represent the flux and model time series, respectively, and $K$ is the covariance matrix.

$K$ is specified by the following hyper-parameters which are allowed to vary as free parameters: the variability amplitude ($A$), the primary variability period ($P_\mathrm{rot}$), the quality factor minus 1/2 for the secondary oscillation ($Q_0$), the difference between the quality factors of the first and second modes ($\Delta Q$), and the fractional amplitude of the secondary mode relative to the primary mode (``mix''). Each of these hyper-parameters except the mixture term was sampled in log space. The white noise amplitude was fixed to 360 ppm, which we estimated from the photometric scatter in-transit and is larger than the scatter out-of-transit due to the presence of likely spot-crossings. 

\subsection{Sampling} \label{subsec:sampling}
An initial maximum a posteriori (MAP) solution was found using \texttt{scipy.optimize.minimize}. From the fit residuals we identified and excluded 10$\sigma$ outliers from further analysis, where $\sigma$ was determined from the root-mean-square of the residuals. This step resulted in the exclusion of 44 points, 6 of which were in-transit observations. A new MAP solution was then derived from the sigma-clipped light curve and used to initialize the parameters sampled with a Markov chain Monte Carlo (MCMC) analysis. The MAP solution was used to initialize the parameters sampled with a Markov chain Monte Carlo (MCMC) analysis. The MCMC sampling was performed using the No U-Turns step-method \citep{hoffman2014}. We ran 4 chains with 500 tuning steps to learn the step size, 9,000 tuning iterations (tuning samples were discarded), a target acceptance of 95\%, and 3000 draws for a final chain length of 12,000 in each parameter. Convergence was assessed using the Gelman-Rubin diagnostic \citep{Gelman:Rubin:1992}, which was below 1.0023 for each parameter. 

The light curve modeling results and derived planetary parameters are summarized in Table~\ref{table:lcresults}. Figure~\ref{fig:lc} shows the full \textit{K2} light curve before and after subtracting the median GP model, along with the phase-folded, whitened photometry and median transit models. Figure~\ref{fig:gp} shows the \textit{K2} data in the regions surrounding individual transits of V1298 Tau along with the GP model predictions for the stellar variability. Finally, three simultaneous transits of planets c and d are shown in greater detail in Figure~\ref{fig:simultaneoustransits}.

\subsection{Separate De-trending and Transit Modeling}
\label{subsec:separate}
We also performed a two step analysis in which the stellar variability was first removed and then the transit model sampling was performed on the flattened light curve. We modeled the stellar variability with a cubic spline with a knot spacing of 6 cadences. Transits and flares were masked from the spline fit using a custom mask that was created by visual inspection of the light curve in a cadence-by-cadence manner. The planet radii derived from the two analyses were consistent at the 1$\sigma$ level.

\subsection{Assessment of Pipeline Sensitivity}
We investigated whether our derived planet parameters were sensitive to our adopted systematics model or stellar variability model. For these purposes, we analyzed photometry from the \texttt{K2SC} \citep{Aigrain2016} pipeline and also experimented with the single Simple Harmonic Oscillator (SHO) GP kernel. In all iterations of our analysis we found the planet radius determinations to be consistent at the $\lesssim$1$\sigma$ level.

\subsection{Modeling the Outer Planet's Single Transit}
We placed a lower limit on the period of planet e of $P > 36$ days from the lack of additional transits in the \textit{K2} light curve. We therefore can not rule out the possibility that V1298 Tau e is in a low-order resonance (of 3:2, 5:3, 2:1, or 3:1) with planet b. 

We modeled the single transit in a similar manner as described above for the other transits. However, following the recommendation of \citet{Kipping:2018}, we imposed a prior on the observed period of:

\begin{equation}
\text{Pr}(P) \propto P^{\alpha-5/3},    
\end{equation}

where we assume that the intrinsic period distribution for an exoplanet in the present regime is proportional to $P^{\alpha}$ and we adopt $\alpha=-2/3$, consistent with the \citet{Burke:etal:2015} analysis of the \textit{Kepler} sample. Additionally, to speed up convergence, we performed our analysis on the flattened light curve described in \S~\ref{subsec:separate}.

As with the transits of V1298 Tau b, there is extra variability in transit which may potentially be due to spot-crossings. Without explicitly modeling or masking this extra variability, we measured a radius for the planet that is about 8\% smaller than Saturn. The results of the single transit fits are presented in Table~\ref{table:lcresults}. 
\section{Discussion} \label{sec:discussion}

\subsection{Estimating Planet Masses}
\label{subsec:masses}
One metric for quantifying the degree to which a planetary system is dynamically packed is the separation in units of mutual Hill radii:

\begin{equation} \label{eq:mutualhill}
    \Delta_H = \frac{a_2-a_1}{R_H} = 2 \frac{\left ( \frac{P_2}{P_1} \right )^{2/3}-1}{\left ( \frac{P_2}{P_1} \right )^{2/3}+1}\left ( \frac{3 M_\star}{M_{P,1}+M_{P,2}} \right )^{1/3}.
\end{equation}

More than 90\% of planet pairs in multiplanet systems have mutual Hill separations of $\Delta_H>10$, and typical spacings range from $\Delta_H\approx10-30$ \citep{Fang:Margot:2013, Fabrycky:etal:2014}. If this pattern holds true for the V1298 Tau system, constraints on the planet masses can be derived by inverting Equation \ref{eq:mutualhill}:

\begin{equation} 
    M_{P,1}+M_{P,2} = 8 \left [ \frac{\left ( \frac{P_2}{P_1} \right )^{2/3}-1}{\left ( \frac{P_2}{P_1} \right )^{2/3}+1} \right ]^3 \frac{3 M_\star}{\Delta_H^3}.
\end{equation}




Using $\Delta_H$ values randomly drawn from a shifted Rayleigh distribution with standard deviation $\sigma=9.5$ \citep{Fang:Margot:2013} we estimated the total mass of the inner two planet pairs from the period ratios. We found median values of $M_{c}+M_{d} = 7^{+21}_{-5} M_\oplus$ and $M_{b}+M_{d} = 29^{+91}_{-20} M_\oplus$, where the uncertainties reflect the 68\% percentile range. We then assumed the inner two planets are of equal mass to find estimates of each of the individual planet masses. Assuming that the V1298 Tau planets will follow the mature exoplanet mass-radius relation of \citet{Chen:Kipping:2017} in the future, we calculated the expected final radii to estimate that the planets might contract by 40--90\% (68\% confidence interval) over the subsequent evolution of the system. These results are not changed significantly when adopting the nonparametric mass-radius relation of \citet{Ning2018}.

\subsection{Eccentricities and Orbit-crossing Constraints}
By requiring the periapsis of each planet to be larger than the semi-major axis of the adjacent interior planet (and the apoapsis to be smaller than the semi-major axis of the adjacent exterior planet), one can further constrain the orbital eccentricities. From the MCMC chain, we enforced the conditions $a_\mathrm{outer}(1-e) > a_\mathrm{inner}$ and $a_\mathrm{inner}(1+e) < a_\mathrm{outer}$ for each adjacent pair of planets. We additionally required the periapsis of the inner planet to be larger than the stellar radius. Before applying these constraints we found 95\% confidence limits to the eccentricities of $e<0.42,0.49,0.40,0.60$ for planets b, c, d, and e, respectively. After applying the orbit-crossing constraints, we find limits of $e<0.43,0.21,0.29,0.57$. Tighter eccentricity constraints might be derived from numerical $N$-body simulations, which we leave to a future work.

\subsection{Proximity to Resonance and Transit-Timing Variations}

Here, we discuss the proximity of planets b, c, and d to mean-motion resonance (MMR) and the implications for additional characterization by transit-timing variations (TTVs). Planets in first order MMR have period ratios of $\frac{P_2}{P_1} = \frac{j}{j-1}$, where $j$ is an integer, and the proximity of a system to resonance is characterized by $\Delta = \frac{P_2}{P_1}\frac{j - 1}{j} - 1$. The c-d pair is close to the 3:2 MMR with $\Delta = 0.25\%$; the d-b pair resides near the 2:1 MMR with $\Delta = -2.6\%$.

\cite{Batygin17} showed that the resonant bandwidth is approximately
\begin{equation}
|\Delta| \lesssim 5
       \left( \frac{j-1}{j^{2/3}} \right) 
       \left( \frac{M_1 + M_2}{M_\star} \right)^{2/3} 
       \simeq 
        0.5 \%
       \left( \frac{M_1 + M_2}{10 M_\oplus} \right)^{2/3} 
       \left( \frac{M_\star}{M_\odot}\right)^{-2/3}
\end{equation}
For the d-b pair, this criterion is satisfied if $M_d + M_b \gtrsim 100$~\mearth, which is consistent with our mass estimates from Section~\ref{subsec:masses}. Planets c and d may be in resonance if $M_c + M_d \gtrsim 4$~\mearth, which is satisfied even for very low-density planets. We emphasize that the above criterion shows that resonant configurations are plausible, but not guaranteed. To confirm a resonant architecture, one must demonstrate libration of resonant angles. Such a confirmation may be possible with future measurements of masses and eccentricities by RVs and/or TTVs along with $N$-body work.

Below, we estimate the magnitude of TTVs assuming non-resonant configurations. If the system is indeed in resonance, the TTVs will depend on the libration width and can be arbitrarily small or an appreciable fraction of the orbital period. This cannot easily be estimated from the available data. For planets close to, but not in, first order MMR, \cite{Lithwick12} showed that the TTV signature is well-described by anti-correlated sinusoids with a ``super-period'' $P^{\prime} = \frac{P_2}{j |\Delta|}$ and amplitudes given by:
\begin{eqnarray}
    V_1 & = & P_1 \frac{\mu_2}{\pi j^{2/3}(j - 1)^{1/3}\Delta} \left(-f - \frac{3}{2} \frac{Z}{\Delta}\right) \\
    V_2 & = & P_2 \frac{\mu_1}{\pi j\Delta} \left(-g - \frac{3}{2} \frac{Z}{\Delta}\right),
\end{eqnarray}
where $Z$ is a linear combination of the planet eccentricities, and $f$ and $g$ are order unity coefficients that depend on $j$ and are given in \cite{Lithwick12}. 


Because $|\Delta_{db}|$ is $\sim10$ times larger than  $|\Delta_{cd}|$ we expect that the c-d interactions will dominate the overall TTV signal, except in the case of extreme mass ratios or eccentricities. For brevity, we provide estimates of the TTVs associated with c-d interactions. However, with sufficient photometric precision, one can also detect the near-resonant TTVs from the d-b interactions as well as higher-order effects such as synodic chopping \citep{Deck15}.

The transits of planets c and d will deviate from a linear ephemeris over a super-period  of 4.5 years. We derive a lower limit to the TTV amplitudes by assuming that the planets have circularized, i.e. $Z = 0$:
\begin{eqnarray}
    |V_c| & \approx & 0.6\, \mathrm{hr} \left( \frac{M_d}{10\, M_\oplus} \right) \\
    |V_d| & \approx & 1.0\, \mathrm{hr} \left( \frac{M_c}{10\, M_\oplus} \right)
\end{eqnarray}
However, the TTV amplitudes may be much larger if the planets have even small eccentricities. \cite{Hadden17} performed an ensemble analysis of 55 near-resonant \textit{Kepler} multiplanet systems and found that $Z$ is typically a few percent, but ranges from 0.0--0.1. When $Z \gg \Delta$:
\begin{eqnarray}
    |V_c| & \approx & 3.5\, \mathrm{hr} \left( \frac{M_d}{10\, M_\oplus} \right)\left( \frac{Z}{0.02} \right) \\
    |V_d| & \approx & 4.6\, \mathrm{hr} \left( \frac{M_c}{10\, M_\oplus} \right)\left( \frac{Z}{0.02} \right).
\end{eqnarray}
Given the large expected TTVs, future transit observations of planets c and d will be particularly valuable for characterizing the masses and eccentricities of these planets.

When planning future observations, however, one must account for the fact that our measured orbital periods were themselves influenced by TTVs. Assuming a strict linear ephemeris, the uncertainty on the time of future transits grows like $\sigma_T = \frac{\sigma_P}{P} \Delta t$, where $\Delta t$ is the time since the {\em K2} epoch. While the periods listed in Table~1 have small fractional uncertainties of $9\times10^{-5}$ and $1 \times 10^{-4}$ respectively, these correspond to the mean periods measured during  {\em K2} observations, which is not the same as the mean periods averaged over a TTV cycle \permean. The mean period, measured over a small section of the TTV cycle may differ from \permean by as much as $\pm 2\pi V \permean/P^\prime$. For even modest TTVs amplitudes of $|V| = $~1~hr, this amounts to a fractional change of $2 \times 10^{-4}$, which is $\sim$2 times larger than those quoted in Table~1. Therefore, future attempts to recover planets c and d should accommodate these expected TTVs.




\subsection{Comparison to the Population of Known Exoplanets}

\begin{figure}
    \centering
    \includegraphics[width=\textwidth]{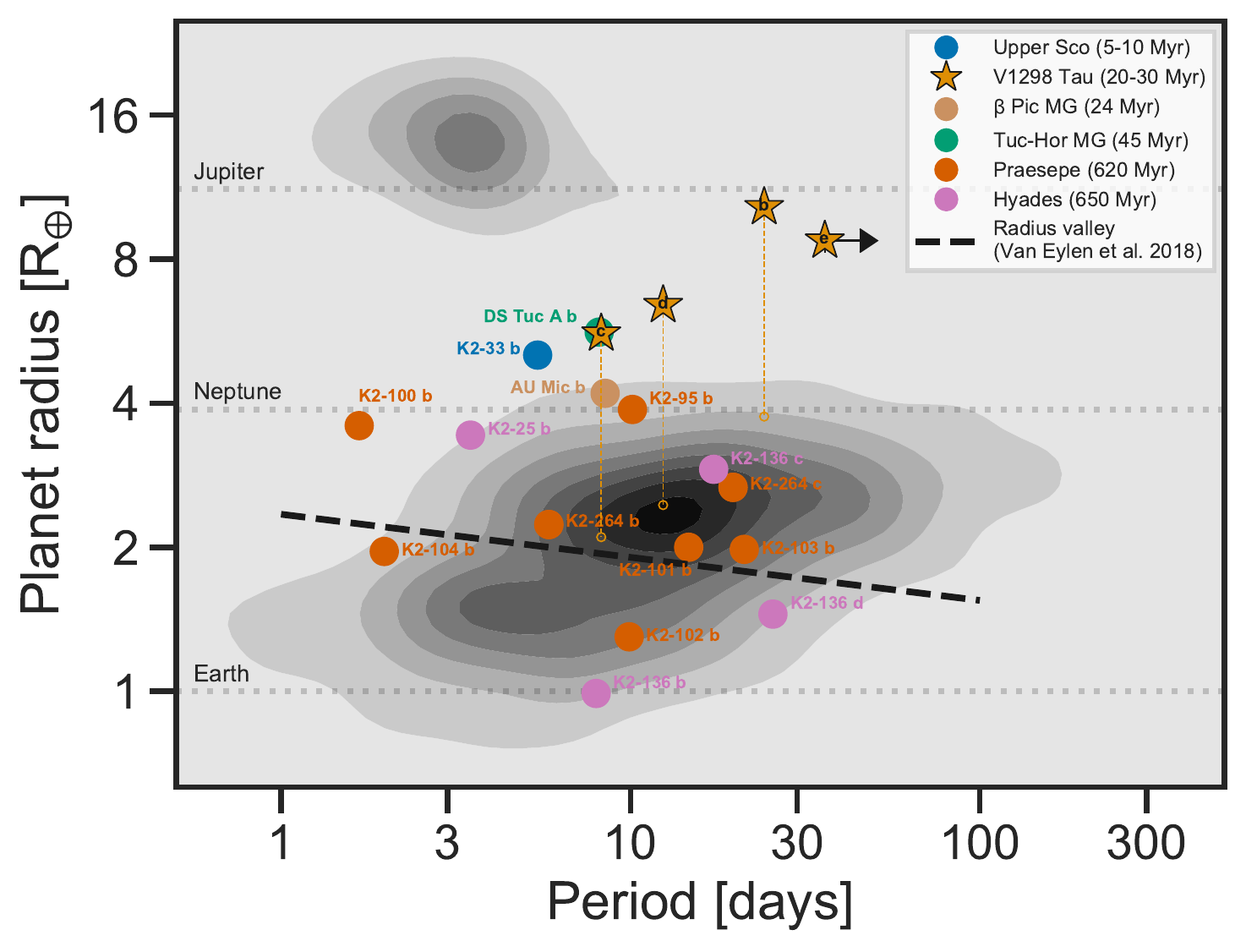}
    \caption{Young transiting exoplanets in the period-radius plane. Contours show a Gaussian kernel density estimate of the distribution of confirmed transiting exoplanets in the period-radius plane. Transiting planets in open clusters or other young stellar associations are indicated by the shaded circles. The planets transiting V1298 Tau are indicated by the gold stars. Vertical dashed lines and open circles in gold represent plausible evolutionary tracks and predicted radii at 5 Gyr, respectively, for planets b, c, \& d based on photo-evaporation models (J. Owen, priv. communication). The slope of the radius valley derived by \citet{vanEylen:etal:2018} is depicted by the black dashed line.}
    \label{fig:pr}
\end{figure}

Using data from the NASA Exoplanet Archive\footnote{\url{https://exoplanetarchive.ipac.caltech.edu}. Accessed on 2019 July 30.} \citep{Akeson:etal:2013} we compared the periods and radii of the planets orbiting V1298 Tau to those of the broader population of known exoplanets. The comparison yields two interesting insights. First, as shown in Figure~\ref{fig:pr}, the planets orbiting V1298 Tau occupy sparsely populated regions of the period-radius plane. In this regard, the V1298 Tau planetary system conforms to the trend of apparently inflated radii which has been noted previously for other young transiting planets \citep{David2016, Mann2016a, Mann2017}. Second, with regards to the number of large transiting planets at small orbital separations, it is clear that the V1298 Tau system is nearly in a class of its known. Of the 539 known multi-transiting systems, only one other star hosts 3 or more planets larger than 5~\rearth with periods $<$300 days: Kepler-51. The Kepler-51 system hosts three planets larger than 7~\rearth inside of 0.51~au, which are the lowest density exoplanets known \citep{Steffen:etal:2013, Masuda:2014}. Perhaps not coincidentally, Kepler-51 has an estimated age of 0.3--0.5 Gyr from gyrochronology \citep{Walkowicz:Basri:2013, Masuda:2014}.

\section{Conclusions} \label{sec:conclusions}
We report the discovery of three additional transiting planets from the \textit{K2} light curve of the young star V1298 Tau. The planets orbiting V1298 Tau join three other recently discovered planets transiting pre-main sequence stars in similarly young associations: K2-33 b in Upper Scorpius \citep{David2016, Mann2016b}, DS Tuc A b in the Tuc-Hor moving group \citep{Newton:etal:2019, Benatti:etal:2019}, and AU Mic b in the $\beta$ Pic moving group (Plavchan et al., submitted). These young planets serve as important benchmarks for planet formation and evolution models. 

Our primary conclusions regarding the planets orbiting V1298 Tau are as follows:

\begin{enumerate}
    \item Assuming typical values for the orbital separations in units of mutual Hill radii, we predict a total mass of 2--28~\mearth for planets c and d and 9--120~\mearth for planets d and b. If confirmed, the low densities implied for these planets indicates (a) the V1298 Tau system may represent a progenitor to the fairly common class of closely-spaced, coplanar, multiplanet systems discovered by \textit{Kepler}, and (b) they are good targets for transmission spectroscopy. 
    \item Estimating individual planet masses and using an exoplanet mass-radius relation calibrated to older systems, we find that the planets orbiting V1298 Tau might contract by 40--90\% during the subsequent evolution of the system.
    \item The proximity of V1298 Tau~c and d to a 3:2 period commensurability suggests that some close-in planets may either form in resonances or evolve into them on timescales of $\lesssim 10^7$~years. One theory for forming resonant chains of planets involves convergent migration of the planets while still embedded in a viscous protoplanetary disk \citep[e.g.][]{Masset:Snellgrove:2001, Snellgrove:etal:2001, Lee:Peale:2002, Cresswell:Nelson:2006, Terquem:Papaloizou:2007} 
    \item The V1298 Tau planetary system constitutes a valuable laboratory for testing photo-evaporation models across a range of incident flux and at a stage when atmospheric loss is expected to be particularly vigorous. Photo-evaporation is expected to play an important role in the evolution of the inner two planets, which may be actively losing atmosphere, but a much lesser role for the outer two planets \citep{OwenWu2013}. Preliminary modeling of the system suggests minimum core masses of 5~\mearth and initial envelope mass fractions of 20\% for each of the three innermost planets, with predicted radii at 5~Gyr of 3.75, 2.1, and 2.45~\rearth for planets b, c, \& d, respectively (J. Owen, priv. communication).
    \item Significant uncertainties remain in the ephemerides of all planetary candidates. The best available ephemeris for V1298 Tau b is presented in a companion paper, which combines \textit{Spitzer} and \textit{K2} transit observations (Livingston et al., in prep.). We advocate for continued monitoring of V1298 Tau to refine ephemerides and search for TTVs. Observing at redder wavelengths, where the star is brighter and the amplitude of stellar variability is lower, is preferred.
\end{enumerate}

\begin{deluxetable*}{rlllll}
\tabletypesize{\scriptsize}
\tablecaption{V1298 Tau light curve modeling results. \label{table:lcresults}}
\tablecolumns{6}
\tablewidth{0pt}
\tablehead{
\colhead{} &
\colhead{} &
\colhead{}
}
\startdata
\hline 
\textit{Star} & \textit{Value} & \textit{Prior} \\
\hline 
$M_\star$ ($M_\odot$) & 1.101$^{+0.049}_{-0.051}$ & $\mathcal{G}$(1.10, 0.05)\\
$R_\star$ ($R_\odot$) & 1.345$^{+0.056}_{-0.051}$ & $\mathcal{G}$(1.305, 0.07)\\
$u_1$ & 0.46$^{+0.22}_{-0.25}$ & $\mathcal{U}$[0,1] in $q_1$\\
$u_2$ & 0.11$^{+0.42}_{-0.34}$ & $\mathcal{U}$[0,1] in $q_2$\\ 
$\langle f \rangle$ (ppt) & 0.00 $\pm$ 0.27  & $\mathcal{G}$(0, 10) \\
$\ln$($A$/ppt) & 4.95$^{+0.66}_{-0.44}$ &  $\mathcal{G}$(190, 5) \\
$P_\mathrm{rot}$ (day) & 2.870 $\pm$ 0.022  &  $\mathcal{G}$($\ln{2.865}$, 5) \\
$\ln$($Q_0$) & 2.42$^{+0.54}_{-0.37}$ & $\mathcal{G}$(1,10)\\
$\Delta Q_0$ & 3.2$^{+1.1}_{-3.6}$ & $\mathcal{G}$(2,10)\\
mix & 0.31$^{+0.34}_{-0.17}$ &  $\mathcal{U}$[0,1]\\
\hline 
\textit{Planets} & \textit{c} & \textit{d} & \textit{b} & \textit{e} \\
\hline
$P$ (days) & 8.24958 $\pm$ 0.00072  & 12.4032 $\pm$ 0.0015 & 24.1396 $\pm$ 0.0018 & 60$^{+60}_{-18}$ \\
$T_{0}$ (BJD-2454833) & 2231.2797 $\pm$ 0.0034 & 2239.3913 $\pm$ 0.0030 & 2234.0488 $\pm$ 0.0018 & 2263.6229 $\pm$ 0.0023 \\
$R_P/R_\star$ & 0.0381 $\pm$ 0.0017 & 0.0436$^{+0.0024}_{-0.0021}$ & 0.0700 $\pm$ 0.0023 & 0.0611$^{+0.0052}_{-0.0037}$ \\
$b$ &  0.34$^{+0.19}_{-0.21}$ & 0.29$^{+0.27}_{-0.20}$ & 0.46$^{+0.13}_{-0.24}$ & 0.52$^{+0.17}_{-0.29}$ \\
$e$ &  $<$0.43 & $<$0.21 & $<$0.29 & $<$0.57 \\
$\omega$ (deg) &  92 $\pm$ 70 & 88 $\pm$ 69 & 85 $\pm$ 72 & 91 $\pm$ 62 \\
$i$ (deg) & 88.49$^{+0.92}_{-0.72}$ & 89.04$^{+0.65}_{-0.73}$ & 89.00$^{+0.46}_{-0.24}$ & 89.40$^{+0.26}_{-0.18}$ \\
$a/R_\star$ & 13.19 $\pm$ 0.55 & 17.31 $\pm$ 0.72 & 27.0 $\pm$ 1.1 & 51$^{+31}_{-11}$\\
$R_P$ ($R_\mathrm{Jup}$) & 0.499$^{+0.032}_{-0.029}$ & 0.572$^{+0.040}_{-0.035}$ & 0.916$^{+0.052}_{-0.047}$ & 0.780$^{+0.075}_{-0.064}$ \\
$R_P$ ($R_\oplus$) &  5.59$^{+0.36}_{-0.32}$ & 6.41$^{+0.45}_{-0.40}$ & 10.27$^{+0.58}_{-0.53}$ & 8.74$^{+0.84}_{-0.72}$ \\
$a$ (au) & 0.0825 $\pm$ 0.0013  & 0.1083 $\pm$ 0.0017  & 0.1688 $\pm$ 0.0026 & 0.308$^{+0.182}_{-0.066}$ \\
$T_{14}$ (hours) & 4.66 $\pm$ 0.12 & 5.59 $\pm$ 0.13  & 6.42 $\pm$ 0.13 & 7.45$^{+0.32}_{-0.25}$ \\
$T_{23}$ (hours) & 4.26 $\pm$ 0.12 & 5.04$^{+0.13}_{-0.18}$ & 5.36$^{+0.14}_{-0.18}$ & 6.24$^{+0.29}_{-0.38}$ \\
$T_\mathrm{eq}$ (K) & 968 $\pm$ 31 & 845 $\pm$ 27 & 677 $\pm$ 22 & 492$^{+66}_{-104}$ \\
$S$ ($S_\oplus$) & 146 $\pm$ 20 & 85 $\pm$ 11 & 35 $\pm$ 5 & 10 $\pm$ 6 \\
\hline 
\textit{Priors} & \textit{c} & \textit{d} & \textit{b} & \textit{e} \\
\hline
$\log{(P/\text{days})}$ & $\mathcal{G}$($\log{8.25}$, 0.1) &  $\mathcal{G}$($\log{12.40}$, 0.1) & $\mathcal{G}$($\log{24.14}$, 0.1)  & -$\frac{7}{3}\log_{10}(P)$, $P:\mathcal{U}(36,1000)$ \\
$T_{0}$ (BJD-2454833) & $\mathcal{G}$(2231.28, 0.25) & $\mathcal{G}$(2239.39, 0.25) & $\mathcal{G}$(2234.05, 0.25) & $\mathcal{G}$(2263.60, 0.25) \\ 
$\log{(R_P/R_\odot)}$ & $\mathcal{G}$(-2.74, 0.2) & $\mathcal{G}$(-2.73, 0.2) & $\mathcal{G}$(-2.36, 0.2) & $\mathcal{G}$(-2.36, 0.2) \\
$b$ & $\mathcal{U}$[0, 1] in $r_1,r_2$ & $\mathcal{U}$[0, 1] in $r_1,r_2$ & $\mathcal{U}$[0, 1] in $r_1,r_2$ & $\mathcal{U}$[0, 1] in $r_1,r_2$ \\
$e$ & $\beta$($a$=0.867, $b$=3.03) & $\beta$($a$=0.867, $b$=3.03) & $\beta$($a$=0.867, $b$=3.03) & $\beta$($a$=0.867, $b$=3.03)
\\
$\omega$ (deg) & $\mathcal{U}$(-180, 180) & $\mathcal{U}$(-180, 180) & $\mathcal{U}$(-180, 180) & $\mathcal{U}$(-180, 180)
\enddata
\tablecomments{Priors are noted for those parameters which were directly sampled. $\mathcal{G}$: Gaussian. $\beta$: Beta distribution. $\mathcal{U}$: Uniform. Quoted transit parameters and uncertainties are medians and 15.87\%, 84.13\% percentiles of the posterior distributions. Quadratic limb darkening coefficients were sampled using the $q_1, q_2$ parametrization of \citet{Kipping:2013}. Joint sampling of impact parameter and radius ratio was performed using $r_1,r_2$ parameterization of \citet{Espinoza2018}. Eccentricity limits are derived from the 95th percentile of the posteriors after applying orbit crossing constraints. Sampling of $\omega$ performed in $\cos\omega,\sin\omega$. Equilibrium temperatures are calculated assuming an albedo of 0.}
\end{deluxetable*}

\acknowledgments \copyright~2019. California Institute of Technology. U.S. Government sponsorship acknowledged. We are grateful to Konstantin Batygin, Elisabeth Newton, Aaron Rizzuto, and Andrew Mann for helpful discussions. T.J.D. and E.E.M. gratefully acknowledge support from the Jet Propulsion Laboratory Exoplanetary Science Initiative. Part of this research was carried out at the Jet Propulsion Laboratory, California Institute of Technology, under a contract with NASA. This paper includes data collected by the {\em Kepler} mission, funded by the NASA Science Mission directorate. This research has made extensive use of the \texttt{exoplanet} documentation and tutorials provided at \url{https://exoplanet.readthedocs.io/en/stable/}.

\vspace{5mm}
\facilities{Kepler}

\software{\texttt{astropy} \citep{exoplanet:astropy13, exoplanet:astropy18},
          \texttt{emcee} \citep{foremanmackey2013},
          \texttt{exoplanet} \citep{exoplanet:exoplanet},
          \texttt{EVEREST 2.0} \citep{Luger2018},
          \texttt{ipython} \citep{ipython},
          \texttt{jupyter} \citep{jupyter},
          \texttt{K2SC} \citep{Aigrain2016},
          \texttt{lightkurve} \citep{lightkurve},
          \texttt{matplotlib} \citep{matplotlib},
          \texttt{numpy} \citep{numpy},
          \texttt{pymc3} \citep{exoplanet:pymc3},
          \texttt{seaborn} \citep{seaborn},
          \texttt{scipy} \citep{scipy},
          \texttt{Starry} \citep{exoplanet:luger18},
          \texttt{theano} \citep{exoplanet:theano}
          }




\end{document}